\documentclass[12pt]{iopart}
\expandafter\let\csname equation*\endcsname\relax
\expandafter\let\csname endequation*\endcsname\relax
\usepackage{amsmath}
\usepackage{dsfont,mathrsfs,bm,xcolor,amsthm}
\usepackage{amstext}
\usepackage{cite}
\usepackage{braket}
\usepackage{soul}
\usepackage[titletoc,title]{appendix}

\theoremstyle{plain}

\theoremstyle{definition}

\begin{document}

\title[]{Reply to the Comment on `The operational foundations of PT-symmetric and quasi-Hermitian quantum theory'}

\author{Abhijeet Alase$^{1}$, Salini Karuvade$^{1}$, Carlo Maria Scandolo$^{2,3}$}

\address{$^1$ Centre for Engineered Quantum Systems, School of Physics, University of Sydney, Sydney, New South Wales 2006,
Australia}
\address{$^2$ Department of Mathematics \& Statistics, University of Calgary, Calgary, Alberta T2N 1N4, Canada}
\address{$^3$  Institute for Quantum Science and Technology, University of Calgary, Calgary, Alberta T2N 1N4, Canada}

\begin{abstract}
This document is our reply to the Comment 
(Miloslav Znojil 2023 \emph{J. Phys. A: Math. Theor.} {\bf 56} 038001) on our recent work titled 
`The operational foundations of PT-symmetric and quasi-Hermitian quantum theory'. 
The original Comment consists of three addenda to our work. 
The first addendum claims that our work is ill-motivated as the motivating question, namely
whether PT-symmetric quantum theory extends the standard quantum theory, was already
answered in the literature. The second addendum points to some missing references in our work, 
and the third addendum suggests what constraints could lead to an extension of standard quantum theory.
In our reply, we explain that the claim in the first addendum is a result of
a misinterpretation of our motivating question. When interpreted correctly,  
the third addendum in the Comment in itself elaborates on why our motivating question 
is interesting and relevant. We also briefly comment on the prospects of an extension of
standard quantum theory along the lines suggested in the third addendum. As our response to 
the second addendum, we explain our rationale behind citing 
certain references while leaving out others.
\end{abstract}

\section*{}

We thank M. Znojil for reading our work with interest and for initiating a constructive exchange. 
Their Comment~\cite{Zno23} on our work~\cite{AKS22} consists of three addenda. 
In the first addendum, it is claimed that our work is ill-motivated as the motivating question, namely
whether PT-symmetric quantum theory extends the standard quantum theory, has been 
answered over 12 years ago. The second addendum points to some missing references. Finally, the third addendum suggests what constraints could lead to an extension of standard quantum theory. In this reply, we comment
on these three addenda. In particular, we emphasize that the author found our work ill-motivated
because they have misinterpreted our motivating question to be alluding to rather restricted definitions 
(which we refer to as settings in the rest of this reply) of PT-symmetric quantum theory, which have been already investigated in the literature.
In fact, the third addendum 
in their Comment in itself elaborates on why our motivating question is interesting and relevant 
even though their addendum is restricted in its scope as it pertains only to non-stationary version of quasi-Hermitian
quantum theory (QHQT). As our response to 
the second addendum, we also explain our rationale behind citing 
certain references while leaving out others. 
Whereas some relevant references were omitted unintentionally, these missing references 
would have in fact served to strengthen the main message of our paper, as we discuss below.

Before diving into the three addenda mentioned above, 
we point out that the correctness of our results was not challenged in Comment~\cite{Zno23}. 
Rather, the motivation behind the work and the interpretation of our results are criticized. 
We also clarify that we do not agree with the author's claim that the main mathematical message of our paper is the 
``compatibility between the three alternative versions
of quantum theory.'' If ``three alternative versions'' 
refer to the three settings we examine in our paper,
then quite clearly they lead to very different physical systems, as stated therein. 
These three settings pertain to three different physical theories: one with PT-symmetric observables, 
one with quasi-Hermitian observables, and one with quasi-Hermitian observables with PT symmetry, given in \S3, \S4 and \S5 of~\cite{AKS22}, respectively.
We show that 
these settings are respectively equivalent to theories with only one trivial state, all states allowed by standard quantum theory and all real quantum states.
The meaning of equivalence in this context is clearly explained in our paper, the idea being that there is a linear bijection between the sets of states and effects of corresponding systems.
Thus, the main message of our paper is conveyed accurately in the abstract of our paper
and is quite different from the one claimed in the Comment under consideration.

In the first addendum provided in \S2 of the Comment, the author claims that the question `whether 
PT-symmetric or quasi Hermitian quantum theories extend standard quantum theory' has been answered more than
12 years ago, and provides some references to back this claim. 
In making this claim, the author is assuming that there exists a certain well-defined theory called  PT-symmetric quantum theory in the literature.
We reiterate, as we did in \S{I} and \S{II} of our paper, that  postulates or axioms of PT-symmetric quantum theory 
have never been written down and/or agreed upon. 
Any work attempting to answer the question given above must
choose a setting that defines PT-symmetric quantum theory. 
For instance, Ref.\ 3 cited in the Comment provides a discussion of various settings that were motivated by PT symmetry and
investigated to form consistent
extensions of standard quantum theory~\cite{Mos10b}.
Some of these settings pertain to the use of indefinite, $CPT$ and other positive-definite inner product spaces, respectively.
The last setting in fact gives rise to QHQT, which is also considered in our paper.
However, Ref.~\cite{Mos10b} does not
preclude the possibility of other settings leading to consistent extensions of standard quantum theory.
Other examples from the literature include Ref.~\cite{Zno15} (Ref.\ 4 in the Comment) and Ref.~\cite{JMCN19}, which consider a setting in which all observables are quasi-Hermitian with respect to a possibly time-dependent metric operator.

In fact, we raised the above question (not for the first time) without 
keeping any particular setting in mind. 
The only guide to our exploration was to base our assumptions on the concepts developed in
the field of PT-symmetry. It is obvious, then, that the question we raised cannot be answered in full generality, as 
any such answer will always depend on the setting under consideration.
Even our work does not answer this
motivating question in its entirety, as
newer settings can lead to different conclusions. 
In the third addendum, the author of the Comment 
seems to have misinterpreted our motivation and work
to be only pertaining to the 
stationary version of 
QHQT. We never made any statement to this effect. This appears to be the root cause of
confusion, which might have motivated a comment from the author in the first place.
Contrary to the claims in the Comment, even the second setting in our paper (see \S{IV} of~\cite{AKS22}), which is the closest to QHQT,
need not be restricted to the stationary case.
We now discuss how the characterization of states in this setting is equally applicable to the non-stationary case. 
Consider a finite-dimensional system with time-dependent metric operator $\eta(t)$ and evolving under
some time-dependent Hamiltonian $H(t)$. 
$H(t)$ is quasi-Hermitian with respect to $\eta(t)$ (see Eq.~(4) in~\cite{AKS22}).
In other words,
$H(t)$ is such that the unitarity of the evolution
is preserved with respect to $\eta(t)$. 
At any time $t$, the states of the 
system are linear functionals on the observables at time $t$, which are also quasi-Hermitian with 
respect to the instantaneous $\eta(t)$. Then, by our analysis in \S{IV}, these states are represented by
$\eta(t)$-density operators. This is in agreement with the literature on QHQT with non-stationary or time-dependent metric operator~\cite{JMCN19}.
Furthermore, by employing the tools developed in~\cite{Mos07,Cro15,KAS22}, one can see that 
for finite-dimensional systems 
a time-dependent metric operator is equivalent to
representing the system in a time-dependent
 Hilbert space.
 Thus, our analysis in \S{IV} dictates that
the non-stationary version of QHQT for finite-dimensional systems 
does not extend the standard quantum theory; rather, similar to the stationary QHQT, it is equivalent 
to representing standard quantum theory in a non-orthogonal basis, albeit a time-dependent one.

Regarding the third addendum given in \S 4 of the Comment,
we welcome the author's optimism about a possible extension of
the standard quantum theory via non-stationary version of QHQT in infinite dimensions.
However, we caution that such an extension 
must also have a consistent interpretation in the general probabilistic theories (GPT) framework, 
unless the notions behind GPTs are being challenged. 
Exploring such a consistent interpretation in the GPT framework is a 
worthwhile exercise in our opinion. It is then apparent that the third and the heaviest addendum
in the Comment in fact serves to elaborate on the validity and the importance of the motivating question 
we raise. We thank M. Znojil for this contribution.

We now address the second addendum in \S 3 of the Comment.
The author has expressed disappointment over the apparent lack
of relevant references on the GPT framework. We cited those references from the GPT literature that
were the most significant for our analysis. We claim neither to provide a historical 
overview of the field nor that the list of references in our paper is exhaustive.
For the same reason, Gudder's seminal work that contributed to the development of GPT is 
not cited. This special issue is dedicated to Gudder's work, and we believe that the 
importance of the contributions by Gudder is beyond dispute, and it does not need
further evidence in the form of citations of works that do not impact our analysis directly.

We thank the author, however, for bringing to our attention two papers~\cite{Pas11,PPR13}, namely Refs.~[18,19] in the Comment, that discuss 
the PT symmetric quantum theory in terms of generalized effect algebras, which are a particular approach to GPTs.
We clarify that these references pertain to setting~2 of our paper (see \S 4), and they were
unintentionally
omitted from our list of references. 
We are apologetic for these omissions. 
Nevertheless, the results of our paper are in complete agreement with 
the results in Refs.~\cite{Pas11,PPR13}.
Therefore, they strengthen the main message of our paper.
On the other hand, it is important to note that settings given in \S {3} and \S {4} of our paper and the approach used to study them are novel and previously unexplored in the literature.

In the introduction of the Comment,
the author of the Comment claims that some of the statements in our paper could mislead the 
reader into believing that PT-symmetric quantum theory is not useful. 
Throughout the paper, we did not make any comment on the utility of PT-symmetric quantum
theory; we restricted our statements to the possibility of an extension of the standard quantum theory,
and backed our claims with rigorous mathematics. We completely 
agree with the author of the 
Comment that PT-symmetric quantum 
theory, in its various forms, has 
engendered new ideas and their realizations 
both in mathematics and physics.

Having finished commenting on all three addenda given in the Comment, we now turn to 
addressing a particular point made therein.
That is, ``the concept
of the ‘extension’ of the existing quantum theory is vague''.
The author implies that the ambiguity in the meaning of extension stems from 
a myriad formulations that are prevalent under the name of standard quantum theory.
We completely disagree with this statement in the Comment. 
Quantum theory as a physical theory has precisely stated postulates, and while it has several
formulations, they are all equivalent.
Indeed, the author correctly notes that various formulations of quantum theory 
``differ dramatically in mathematical and conceptual overview, yet each
one makes identical predictions for all experimental results''. 
Thus, what quantum theory is is not vague.
On the contrary, it provides a concrete prescription for predicting observable 
quantities irrespective of the formalism
being used. 
Perhaps the author means to argue that, while standard quantum theory is not vague,
the meaning of `extending' this theory is unclear. A point of confusion here is the meaning of the
word `extension' itself. The author of the Comment, on multiple occasions therein, 
has used this term to mean both an `interpretation of a physical theory' 
(e.g.\ ``benchmark models still wait for a ‘meaningful extension’ of their fully consistent
GPT interpretation'') and,
more vaguely, to mean a new theory that subsumes the old theory. As for the former context,
the word `interpretation' is more appropriate and accepted in the community.
What we mean by an extension of the standard quantum theory is another theory
that consists of new states that are not in bijection
with the states of the former (cf.\ Def.~2.10 in~\cite{AKS22}). 
In other words, if any state in the new theory does not have a counterpart in
the standard quantum theory, then that theory would be an extension of the latter.
We therefore disagree with the author's statement that ``the concept
of the `extension' of the existing quantum theory is vague''.
We however agree that one formulation of quantum theory may turn out to be 
more suited than others in the quest for such extensions.

In conclusion, much of the content of the Comment under consideration appears to have stemmed from a misinterpretation 
of the motivating question we pose in our original paper. Ironically, the final addendum in the Comment 
reiterates the validity and importance of the motivating question pursued in our work. While we disagree
with the author of the Comment on a couple of points, none of the criticisms provided therein 
has any implications on the correctness 
of our results. 

\paragraph{Acknowledgements}
CMS acknowledges the support of
the Natural Sciences and Engineering Research Council
of Canada (NSERC) through the Discovery Grant “The
power of quantum resources” RGPIN-2022-03025 and the
Discovery Launch Supplement DGECR-2022-00119.

\section*{References}
\bibliographystyle{iopart-num}
\bibliography{ref}

\end{document}